\documentclass[a4paper,twoside]{article}

\usepackage{epsfig}
\usepackage{calc}
\usepackage{amssymb}
\usepackage{amstext}
\usepackage{amsmath}
\usepackage{multicol}
\usepackage{pslatex}
\usepackage[normalem]{ulem}

\usepackage[utf8]{inputenc}
\usepackage[T1]{fontenc}    % use 8-bit T1 fonts
\usepackage{hyperref}       % hyperlinks
\usepackage{url}            % simple URL typesetting
\usepackage{booktabs}       % professional-quality tables
\usepackage{amsfonts}       % blackboard math symbols
\usepackage{nicefrac}       % compact symbols for 1/2, etc.
\usepackage{microtype}      % microtypography
\usepackage{comment}

\usepackage{graphicx}
\usepackage{doi}
\usepackage{amsfonts}
\usepackage{xcolor}

\definecolor{keptblue}{HTML}{1F4E9C}\definecolor{neglgrey}{HTML}{9A9A9A}\definecolor{optorange}{HTML}{D95F02}

\usepackage{physics}
\usepackage{makecell}
\usepackage{multirow}
\usepackage{enumitem}
\usepackage{duckuments}

\usepackage{tikz}

\usetikzlibrary{positioning}

\usepackage{physics}

\usepackage{mathtools} %for \coloneqq :=
\usepackage{array}
\usepackage{wrapfig} %text next to figure
\usepackage{svg}
\usepackage[capitalise]{cleveref}
\usepackage[natbib=true, sorting=none, maxcitenames=1, style=numeric-comp]{biblatex}

\usepackage[colorinlistoftodos,prependcaption,textsize=tiny]{todonotes}
\usepackage{subfiles}
\usepackage[ruled,vlined]{algorithm2e}

\usepackage{xcolor}

\newif\ifdraftcomments
\draftcommentstrue % Use \draftcommentsfalse to hide comments

\usepackage{SCITEPRESS}     % Please add other packages that you may need BEFORE the SCITEPRESS.sty package.

\begin{document}

\title{A Realistic End-to-End Runtime Crossover Estimation for Quantum-Enhanced Training of Large-Scale Machine Learning Models}
%\title{Runtime Crossover of Carleman-Linearized Quantum Training Against Stochastic Gradient Descent}

\author{\authorname{
Maximilian Alexander Wolf\sup{1, 2, 3}, 
Alona Sakhnenko\sup{1, 2}, %\orcidAuthor{0000-0002-3873-710X}
Florian Geissler\sup{1}, %\orcidAuthor{/0000-0002-9776-9603} 
Andrea Matic\sup{1}, \\
Jeanette Miriam Lorenz\sup{1,3} %\orcidAuthor{0000-0001-6530-1873}
}
\affiliation{\sup{1}Fraunhofer Institute for Cognitive Systems IKS,  Munich, Germany}
\affiliation{\sup{2}Technical University of Munich, Munich, Germany}
\affiliation{\sup{3}Ludwig-Maximilian University, Munich, Germany}
\email{jeanette.miriam.lorenz@iks.fraunhofer.de}
}


% \abstract{The abstract should summarize the contents of the paper and should contain at least 70 and at most 200 words. The text must be set to 9-point font size.}
%\abstract{Quantum linear-system solvers promise a polylogarithmic dependence on the problem size and Neural-network training can be formulated by Carleman-linearizing the gradient step and solving the lifted history system on a fault-tolerant quantum computer. We price one download--upload window of this pipeline using a fault-tolerant resource model, including the oracle queries and a sparse readout. Three comparisons have been made each with a different complexity level. At a classical rate of $10^{9}\,\mathrm{FLOP/s}$, the solver alone beats a CGNE solve of the lifted system beyond $XXXXXXX$ parameters and the full solve with readout beyond $XXX$. The complete window beats stochastic gradient descent only beyond $XXXXXX$ parameters, for a small number of retained weights, and is $XXXXX$ times slower for a ResNet with $7.5\times10^{6}$ parameters.} 

\abstract{Machine Learning (ML) is one promising application domain for Quantum Computing (QC) speedups, but asymptotic complexity considerations often mask the actual engineering costs of actual end-to-end workflows. For the use case of training a linearized Neural Network with a Quantum Linear Solver, a provably favorable quantum scaling has been suggested. We present realistic resource estimates for this problem, accounting for data access, computation, and readout. Our analysis examines three increasingly complex benchmarking scenarios, under distinct sparsity-scaling assumptions, thereby enabling progressively more realistic resource estimates. We identify crossover regimes relative to classical methods and find that crossover points emerge only for very large problem sizes, long computation times, and highly restrictive output requirements. This work provides a blueprint for evaluating quantum-enhanced ML algorithms under realistic hardware constraints and help identifying the bottlenecks that must be overcome to approach industrial quantum utility.}

\onecolumn \maketitle \normalsize \setcounter{footnote}{0} \vfill

\section{\uppercase{Introduction}}
\label{sec:introduction}

Training modern machine learning (ML) models and, in particular, large language models (LLMs) has become one of the most computationally demanding workloads in artificial intelligence. Publicly reported training runs illustrate this scale: the 65-billion-parameter LLaMA model was trained for approximately 1.032 million A100 GPU-hours (21 days on 2,048 A100 GPUs), while the full training run of DeepSeek-V3 is reported to require 2.788 million H800 GPU-hours \cite{Touvron_2023,DeepSeekAI_2024}. The resulting financial cost and time effort demand fundamentally more efficient training methods.

Fault-tolerant quantum computing may provide one such route. A growing body of theoretical work analyzes fault-tolerant quantum algorithms with provable asymptotic speedups for computationally demanding ML tasks \cite{Schuld2021}, including training \cite{Zhao_2019,Liu_2024}, formal verification \cite{10821425}, and sampling \cite{Low_2014}. \citet{Liu_2024} provide, to our knowledge, the first theoretical demonstration that quantum computing can exponentially accelerate the full training routine. This approach reformulates stochastic gradient descent (SGD) as a dissipative quadratic ordinary differential equation (ODE), applies Carleman linearization, and solves the resulting lifted-history system using a fault-tolerant quantum linear solver (QLS) algorithm such as HHL. Under its asymptotic assumptions (and certain simplifications we discuss in detail in this article), this construction promises a complexity that scales only polylogarithmically with the number of neural network (NN) parameters, in contrast to a linear scaling achieved by the best known classical algorithm.

However, an asymptotic speedup alone does not establish a practical quantum advantage. The analysis of \citet{Liu_2024} focuses on complexity arguments and does not provide an end-to-end quantitative estimate of the pipeline's resource requirements. In particular, the costs of oracle queries, logical operations, tomography and classical preparation can substantially change the comparison with classical methods. Consequently, the crossover points (the problem scales at which the total cost of the quantum pipeline becomes favorable relative to classical baselines) remain undetermined.

Our work addresses this gap and provides a full end‑to‑end resource estimate that explicitly accounts for these components. We compare this complete quantum pipeline against both the direct classical counterpart (ODE solver) and a non-linearized standard training approach, namely SGD. 
This way, we consider three benchmarking scenarios of increasing realism, each based on different assumptions about sparsity scaling and implementation costs. 
By accounting for all relevant processing steps, our resource model is able to establish practically relevant crossover points between CPU and QPU\footnote{Quantum Processing Unit (QPU)} scenarios in terms of wall-clock runtime and NN parameter size.

Our results show that a realistic quantum speedup requires several demanding conditions to be satisfied simultaneously, particularly concerning favorable sparsity scaling and low gradient construction overheads. On top of this challenge, the crossover points identified under our benchmark assumptions lie in resource regimes of very large compute times ($>10^{19}$ seconds) and NN model sizes ($>10^{20}$ parameters) for the full training pipeline. 
Based on those findings, we discuss, if such assumptions could be feasible for realistic training pipelines and near-future fault-tolerant quantum hardware. Our analysis therefore translates an asymptotic proposal into quantitative conditions for practical quantum advantage.

\section{\uppercase{Related work}}\label{sec:related}
\citet{Liu_2024} provide an explicit claim in which a QLS is used in certain steps of NN training. It reformulates SGD as a dissipative linear ordinary differential equation (ODE) and determine a complexity of $\mathcal{O}\bigl(T \times \mathrm{poly}(\log d, 1/\epsilon)\bigr)$ in the model size $d$, the number of training steps $T$ and allowed error $\epsilon$. The proposed pipeline requires dissipativity, sparsity, and a sufficiently small learning rate. Our work extends \cite{Liu_2024} by pinpointing the regime in terms of $d$, in which the proposed pipeline can surpass their (best) classical alternatives in terms of runtime.

With the recent advances in quantum hardware and a slow transition into Early Fault-Tolerant (EFT) era \cite{PRXQuantum.5.020101}, more scientific inquiry has been directed towards practicality of (Fully) Fault-Tolerant (FFT) algorithms. \citet{babbush2021focus} show that \textit{quadratic speedups} are unlikely to survive the required Quantum Error Correction (QEC) overhead when the classical alternative can be parallelized. In an end-to-end analysis of Quantum Interior-Point methods, \citet{dalzell2023} find that large prefactors, poor conditioning, and extracting the final state via tomography eliminate the advantage. \citet{hoefler2023} study the crossover for a wide rage of FFT algorithms and conclude that practical quantum advantage is not plausible for a many proposed FFT application without signification algorithmic improvements, particularly enforcing light-weight input and output subroutines and striving for \textit{superquadratic speedups}. It is therefore essential to perform an end-to-end estimation of the FFT application to determine the conditions for a practical quantum advantage, which is the goal of this work.

Building on this, here we focus on QLS application, for which \citet{scherer2017} provided one of the earliest concrete resource estimates. An explicit end-to-end runtime estimate of the FFT algorithm against a classical ODE method was performed by \citet{tu2025identifyingpossiblefaulttolerantadvantage}. Their analysis, however, considers a generic use-case, treats the condition number and sparsity as free parameters, and omits the costs of state preparation, oracle implementation, and output recovery. In contrast, our work builds on \cite{tu2025identifyingpossiblefaulttolerantadvantage} but tailors the estimate to a specific use-case from \cite{Liu_2024}, computing the full end-to-end cost and comparing it against both the classical ODE method and SGD costs.

The assumptions required for asymptotic speedups in quantum linear solvers have been discussed extensively. \citet{Aaronson_2015} discusses the input and output, sparsity, and condition-number assumptions that determine whether the exponential speedup of HHL \cite{harrow2009} can be retained.
\cite{DUAN2020126595}

%Most algorithmic studies state their assumptions without assigning costs to them, such as the circuit execution time. Resource estimates, in contrast, consider generic subroutines under assumed condition numbers and sparsities, but they do not derive these variables from a concrete application and tend to omit the pre- and post-processing steps surrounding the solver. This paper estimates these costs and the resulting crossover point if a full run of the pipeline is compared to classic algortihms for solving ODEs and for training neural networks.

\section{\uppercase{Background}}\label{sec:background}
\begin{table}[h]
\centering
\begin{minipage}{\columnwidth}
\centering
\footnotesize
\setlength{\tabcolsep}{3pt}
\renewcommand{\arraystretch}{0.92}

\begin{tabular}{@{}>{\raggedright\arraybackslash}p{0.19\linewidth}>
{\raggedright\arraybackslash}p{0.74\linewidth}@{}}
\toprule
Symbol & Meaning \\
\midrule
$d$ & Model parameter or state dimension. \\
$K$ & Carleman truncation order, at which a polynomial update of degree $q$ is cut off.\\
%$q$ & Degree of the polynomial update. \\
$T$ & Number of training-update intervals. \\
$\Delta_K$ & Lifted dimension $\sum_{j=1}^{K}d^j$. \\
$N_h$ & Matrix dimension with history, $(T+1)\Delta_K$. For the  one-sparse Hamiltonian-simulation (HS1), we use $N_{\mathrm{HS1}}$. \\
$S_{\max}$ & Maximum number of nonzero terms in one original update equation. \\
$s$ & Sparse-access complexity of a history matrix, e.g., $s_M$ for matrix $M$ or  $s_{\mathrm{HS1}}$ for the HS1 resource model. \\
$\kappa_{\mathrm{HS1}}$ & Condition-number input used in the HS1 resource model. \\
$\epsilon$ & Generic target error tolerance used in asymptotic complexity statements. $\epsilon_j(t),\epsilon(t)$ are the blockwise and stacked Carleman-truncation errors. We use
$\epsilon_{\rm HHL},\epsilon_{\mathrm{HS1}}$ for the quantum-state error and the corresponding HS1 resource-model input. $\epsilon_{\mathrm{tom}}$ is the target error of the tomography/readout stage, and $\epsilon_{\mathrm{CGNE}}$ the relative error of the classical CGNE solution.\\
% $\epsilon_{\mathrm{tom}}$ & Target error of the tomography/readout stage. \\
% $\epsilon_j(t),\epsilon(t)$ are the blockwise and stacked Carleman-truncation errors. \\
$B$ & SGD batch size. \\
$r$ & Number of nonzero entries in the parameter vector when preparing the initial state.\\
$r_{\rm out}$ & Support of the normalized terminal output amplitudes, i.e., the $r_{out}$ largest amplitudes.\\
$\eta$ & Learning rate. \\
\bottomrule
\end{tabular}

\end{minipage}
\caption{Overview of notation symbols used in this article.}
\end{table}

% \begin{table}[t]
% \centering
% \renewcommand{\arraystretch}{0.92}

% \end{table}
\subsection{Classical Pipeline}\label{sec:classical_pipeline}

This section first describes the classical workflow that approximates the gradient descent training process as the solution of a linear equation, updating all parameters at once.

\subsubsection{Carleman Linearization}

%We follow the notation of \cite{Liu_2021,Liu_2024}.\\
\textbf{Gradient-descent dynamics.}
Let \(L_A\) be a loss function for a model with parameter vector
\(\theta=(\theta_\mu)\in\mathbb{R}^{d}\).  The starting point is the
discrete gradient-descent update
\begin{equation}
\label{eq:discrete_gradient_descent}
\theta_\mu(t+1)
=
\theta_\mu(t)
-\eta
\left.
\frac{dL_A}{d\theta_\mu}
\right|_{\theta(t)} .
\end{equation}
Here, \(\eta>0\) is the learning rate.  Stochastic gradient descent (SGD)
can be represented by allowing coefficients in the update rule in a training step to depend only on a single, randomly selected sample or a small batch of examples from the dataset, instead of the entire dataset. We denote as base SGD pipeline a workflow that uses the update in Eq.~(\ref{eq:discrete_gradient_descent}) for the time steps $t=0,\ldots,T-1$.\\
A Carleman-linearized pipeline \cite{Liu_2021,Liu_2024} aims to map the above updates to a linear system of the form $M_h x=b$, where $x$ is some representation of the model parameters $\theta$ for the \textit{entire} time sequence of $T$ steps. Hence, solving the resulting linear system is equivalent to solving the full gradient descent dynamics at once. Since \(L_A\) is in general a highly non-linear function of $\theta$, this linearization comes at the cost of truncation errors, which are controlled under certain conditions, e.g., related to the dissipation properties of the Carleman matrix $A$ (see Sec.~\ref{sec:carleman_requirements}).

\textbf{Polynomial form of the update.}
Let us define
$u(t) \coloneqq \left(\theta_\mu\right)(t)$,
$\delta o \coloneqq o(t+1)-o(t)$.
If the gradient update is a polynomial of degree \(q\) in the
parameters, or is approximated by such a polynomial, then
Eq.~\eqref{eq:discrete_gradient_descent} can be written as
\begin{equation}
\begin{aligned}
\delta u
&= u(t+1)-u(t) \\
&= \sum_{\ell=0}^{q} F_\ell\,u^{\otimes \ell}(t) \\
&= F_q\,u^{\otimes q}(t)
 + \cdots
 + F_2\,u^{\otimes 2}(t)
 + F_1\,u(t)
 + F_0 .
\end{aligned}
\label{eq:polynomial-update}
\end{equation}
Here, \(u^{\otimes \ell}\) denotes the \(\ell\)-fold tensor
(Kronecker) product of \(u\) with itself, i.e.,
\(u^{\otimes \ell} = \bigotimes_{r=1}^{\ell} u \in \mathbb{R}^{{d}^{\ell}}\),
with components
\(\left(u^{\otimes \ell}\right)_{i_1,\ldots,i_\ell}
= u_{i_1}\cdots u_{i_\ell}\); For example,
$u^{\otimes 0}=1$ by convention, $u^{\otimes 1}=u$, while
the second tensor power is
$u^{\otimes 2}=u\otimes u=
\left(u_1^2,u_1u_2,\ldots,u_1u_{{d}},
u_2u_1,\ldots,u_2u_{{d}},\ldots,
u_{{d}}u_1,\ldots,u_{{d}}^2\right)^{\mathsf T}$.
% %Equivalently,
% \(\left(u^{\otimes 2}\right)_{i,j}=u_i u_j\) for each ordered pair
% \(\left(i,j\right)\). Although \(u_i u_j=u_j u_i\) as scalar values,
% these occur in separate indexed entries of the Kronecker product.
%
The coefficient tensors \(F_\ell\) contain the polynomial
coefficients of the gradient update.
%, including the factor \(-\eta\).
%In particular, \(F_0\) is the constant part of the update and may be
%time dependent when it represents an additive stochastic contribution.
Explicitly, they are the Taylor coefficients of the gradient-update map expanded around a reference point $\theta_\star$ (say $u:=\theta-\theta_\star$):
\[
(F_\ell)_{\mu;\nu_1\ldots\nu_\ell}
=
-\frac{\eta}{\ell!}
\left.
\frac{\partial^{\ell+1}L_A(\theta)}
{\partial\theta_\mu\partial\theta_{\nu_1}\cdots\partial\theta_{\nu_\ell}}
\right|_{\theta=\theta_\star}.
\]
The leading terms are the first, and second derivates, respectively, 
$(F_0)_\mu
=
-\eta
\left.
\frac{\partial L_A(\theta)}{\partial \theta_\mu}
\right|_{\theta=\theta_\star}$, and
$(F_1)_{\mu\nu}
=
-\eta
\left.
\frac{\partial^2 L_A(\theta)}
{\partial \theta_\mu\,\partial \theta_\nu}
\right|_{\theta=\theta_\star} =-\eta H(\theta_\star)$, where $H$ is the loss \textit{Hessian}.
The tensors $F_{\ell\geq2}$ contain higher derivatives of the loss, which in practice are typically small.

\textbf{Carleman lift.}
The nonlinearity in Eq.~\eqref{eq:polynomial-update} arises from the
tensor powers \(u^{\otimes \ell}\). Carleman linearization promotes
each tensor power to an independent coordinate. The resulting lifted
system is
\begin{equation}
\delta
\left[
\begin{array}{c}
u \\[-2pt]
u^{\otimes 2} \\[-2pt]
u^{\otimes 3} \\[-2pt]
\vdots \\[-2pt]
u^{\otimes (K-1)} \\[-2pt]
u^{\otimes K} \\[-2pt]
\vdots
\end{array}
\right]
=
A
\left[
\begin{array}{c}
u \\[-2pt]
u^{\otimes 2} \\[-2pt]
u^{\otimes 3} \\[-2pt]
\vdots \\[-2pt]
u^{\otimes (K-1)} \\[-2pt]
u^{\otimes K} \\[-2pt]
\vdots
\end{array}
\right]
+
\left[
\begin{array}{c}
F_0 \\[-2pt]
0 \\[-2pt]
0 \\[-2pt]
\vdots \\[-2pt]
0 \\[-2pt]
0 \\[-2pt]
\vdots
\end{array}
\right].
\label{eq:carleman-lift}
\end{equation}

In Eq.~\eqref{eq:carleman-lift}, \(A\) is the \textit{quantum Carleman matrix} (QCM). For the
\(j\)-th block equation, Eq.~\eqref{eq:polynomial-update} is inserted
into each of the \(j\) factors of \(u^{\otimes j}\). This produces
linear couplings from tensor order \(j\) to tensor orders
\(j-1,\ldots,j+q-1\).

The lifted system is formally infinite, so in practical setups it must be truncated at tensor
order \({K}\). The truncated lifted vector is defined as
\begin{equation*}
\begin{aligned}
\hat{y}
&\coloneqq
\left(
u,
u^{\otimes 2},
\ldots,
u^{\otimes K}
\right)^{\mathsf{T}}, 
\end{aligned}
\end{equation*}
while the dimension of the truncated QCM becomes
\begin{equation}
\begin{aligned}
\Delta_K
&\coloneqq
\sum_{j=1}^{K}d^j
=
d+d^2+\cdots+d^K
=
\frac{d^{K+1}-d}{d-1}.
\end{aligned}
\end{equation}

% The omitted higher-order tensor powers produce the Carleman
% truncation error. In the strictly discrete setting, products of
% one-step increments also generate corrections scaling as
% \(\mathcal{O}\!\left(\eta^2 F_2\right)
% +\mathcal{O}\!\left(\eta^3 F_3\right)+\cdots\).
% The small-learning-rate assumption ensures that these corrections
% remain controlled.

The omitted higher-order tensor powers produce the Carleman
truncation error. In the strictly discrete setting, products of
one-step increments also generate corrections scaling as
\(\mathcal{O}\!\left(\eta^2\right)
+\mathcal{O}\!\left(\eta^3\right)+\cdots\) \cite{Liu_2024}.
The small-learning-rate assumption ensures that these corrections
remain controlled.

\textbf{Linear training system.}
Writing the truncated version of Eq.~\eqref{eq:carleman-lift} as
\begin{align*}
\delta \hat{y}(t) &= A\hat{y}(t)+b, \\
\hat{y}(t+1)-\left(I+A\right)\hat{y}(t) &= b,
\end{align*}
the \(T+1\) time slices can be stacked into one block
lower-bidiagonal linear system:
% \begin{equation}
% \left[
% \begin{array}{cccc}
% I & 0 & \cdots & 0 \\
% -\left(I+A\right) & I & \ddots & \vdots \\
% 0 & \ddots & \ddots & 0 \\
% 0 & \cdots & -\left(I+A\right) & I
% \end{array}
% \right]
% \!
% \left[
% \begin{array}{c}
% \hat{y}(0) \\
% \hat{y}(1) \\
% \vdots \\
% \hat{y}(T)
% \end{array}
% \right]
% =
% \left[
% \begin{array}{c}
% \hat{y}_{\mathrm{in}} \\
% b \\
% \vdots \\
% b
% \end{array}
% \right].
% \label{eq:block-linear-system}
% \end{equation}
\begin{equation}
\begingroup
\setlength{\arraycolsep}{2pt}% default is 5pt
\begin{bmatrix}
 I & 0 & \cdots & 0 \\
 -(I+A) & I & \ddots & \vdots \\
 0 & \ddots & \ddots & 0 \\
 0 & \cdots & -(I+A) & I
\end{bmatrix}
\!
\begin{bmatrix}
 \hat{y}(0) \\
 \hat{y}(1) \\
 \vdots \\
 \hat{y}(T)
\end{bmatrix}
=
\begin{bmatrix}
 \hat{y}_{\mathrm{in}} \\
 b \\
 \vdots \\
 b
\end{bmatrix}.
\endgroup
\label{eq:block-linear-system}
\end{equation}

Here,

\begin{align}
\hat{y}_{\mathrm{in}}
&=
\left(
u_{\mathrm{in}},
u_{\mathrm{in}}^{\otimes 2},
\ldots,
u_{\mathrm{in}}^{\otimes K}
\right)^{\mathsf{T}}, \notag \\
b &= \left(F_0, 0, 0, \ldots\right)^{\mathsf{T}}, \notag
\end{align}

The first block row fixes the initial condition, while each later
block row implements one lifted gradient-descent step. For a
time-dependent stochastic update, the corresponding \(A(t)\) and
\(b(t)\) are used in the block row for time step \(t\).
We have therefore transformed the problem into a linear system of the form 
\begin{equation}
    M_h x=b,
\label{eq:linear_equation_generic}
\end{equation}
with ${M_h\in\mathbb{C}^{N_h\times N_h}}$, and ${N_h=(T+1)\Delta_K}$.
An HHL-type quantum linear-system solver can prepare an amplitude
encoding of the solution to Eq.~\eqref{eq:block-linear-system}. Reading out the
first block of ${d}$ elements in \(\hat{y}(T)\) eventually gives the desired parameter vector,
$u(T)=\left(\theta_\mu\right)(T)$.

\subsection{Quantum-enhanced pipeline}
\label{sec:quantum_pipeline}

The quantum pipeline refers to the application of a quantum algorithm, here the HHL \cite{harrow2009} solver used in \cite{Liu_2024} and adopted in our work, to produce a solution to this equation.

\paragraph{HHL fundamentals.}
In Eq.~\eqref{eq:linear_equation_generic}, ${M_h}$ is a non-Hermitian history matrix that requires a solver-compatible embedding, assumed structured-access parameter ${s_M}$, and condition number
${\kappa_M}$. HHL prepares a quantum state proportional to the solution
${x=M_h^{-1}b}$: a clock register is used for phase estimation to access the
eigenvalues of ${M_h}$, which a controlled rotation then inverts in the main
register holding $\ket{b}$. Provided ${M_h}$ has structured access parameter ${s_M}$ and is efficiently
simulable, this state is obtained to precision ${\epsilon_{\rm HHL}}$ in time
${O\!\big(\log(N_h)\,s_M^2\kappa_M^2/\epsilon_{\rm HHL}\big)}$ \cite{harrow2009}, an
exponential speedup in $N_h$ over classical solvers such as the conjugate
gradient method.

\citet{Liu_2024} describe their solver as a variant of the HHL algorithm
\cite{harrow2009}, whose general form they take from \cite{costa2022}: given
a sparse matrix oracle for ${M_h}$ and an oracle for $\ket{b}$, it produces the
state ${\ket{M_h^{-1}b}}$ to error ${\epsilon_{\rm HHL}}$ using gate complexity
${O\big(s_M\kappa_M\cdot\mathrm{polylog}(N_h/\epsilon_{\rm HHL})\big)}$, where
${\kappa_M=\norm{M_h}\norm{M_h^{-1}}}$.

If the Carleman matrix $A$ is fully dissipative, i.e.\ $\norm{I+A}<1$, they
show that the condition number is linearly bounded by the number of
iterations, ${\kappa_M\le\bar\kappa_M:=2(T+1)}$, so that, together with the ${O(s_Mq)}$ cost of the
sparse matrix oracle for $A$, the resulting gate complexity is
${O(s_MqT\cdot\mathrm{polylog}(N_h/\epsilon_{\rm HHL}))}$.

This asymptotic statement carries no explicit constants. Query-count constants for such solvers exist~\cite{jennings2023}, but no fault-tolerant, surface-code-level resource estimate has been published for the solver of \cite{costa2022}. The cost model of Section~\ref{sec:method} therefore prices the phase-estimation solver of \cite{tu2025identifyingpossiblefaulttolerantadvantage}, which is the only quantum linear-system algorithm with a published estimate of that kind. That solver scales as $\kappa_M^{2}/\epsilon_{\rm HHL}^{2}$ rather than $\kappa_M\log(1/\epsilon_{\rm HHL})$, so our estimates are conservative with respect to the algorithm Liu et al.\ invoke. 
% We quantify the difference in Section~\ref{sec:results}: at the reference point it is about four orders of magnitude, which makes the solver choice the single largest lever in the model.

If instead $A$ contains divergent modes, i.e.\ eigenvalues with
$\abs{1+a_\mu}\ge1$, this linear bound on ${\kappa_M}$ no longer holds, and
\cite{Liu_2024} instead deploys a truncated variant that reuses the
ill-conditioned-eigenspace filtering technique from the discussion section of
\cite{harrow2009}: an auxiliary qubit separates eigenspaces with
$\abs{\lambda}\ge 1/\kappa_a$ from those with $\abs{\lambda}<1/\kappa_a$, for a
chosen effective condition number $\kappa_a$, and the state is postselected
onto the well-conditioned branch. Choosing $\kappa_a=2(T+1)$ gives the gate
complexity ${O(s_M^2q^2T^2\cdot\mathrm{polylog}(N_h/\epsilon_{\rm HHL}))}$.

%\begin{figure}[h]
%\centering
%    \includegraphics[width=0.5\textwidth]{figures/hhl.pdf}
%    \caption{Vanilla HHL architecture that solves $A x = b$, where $S_1$ is state preparation of $b$, $S_2$ is QPE (that also encodes $A$ as a unitary), $S_3,*$ is eigenvalue inversion. The blue gate color symbolizes processes, whose efficiency is crucial for the efficiency of the entire HHL.}
%    \label{fig:hhl_circuit}
%\end{figure}

\paragraph{Oracle problem/input encoding.}
The gate complexities discussed above are stated relative to two
oracles: a sparse matrix oracle for the history matrix $M_h$, and an oracle
for $\lvert b \rangle$.
Efficient preparation of the initial state is essential for applying HHL. In
general, this may require QRAM~\cite{QRAM}, whose large-scale realization remains
challenging. Under sparse training, this requirement is relaxed: since
$\theta(0)$ retains bounded support independent of the model size, it can
instead be loaded without QRAM, using sparse-state-preparation circuits that
load an $r$-sparse vector of dimension $n$ with $O(r^2 \log r \log n)$
classical preprocessing and a quantum circuit of depth $O(r \log n)$ composed
only of single-qubit and CNOT gates~\cite{gleinig2022}. 
%For dense weight vectors, this route is unavailable, and loading $\theta(0)$ requires QRAM.
For the matrix oracle, Liu et al.\ relate the sparsity of the QCM $A$ to the
underlying machine-learning architecture: pruning the network so as to reduce
the number of non-zero terms in each gradient-update function correspondingly
bounds the number of non-zero entries per row of the truncated $A$. This
provides the sparsity input required for the matrix oracle.

\paragraph{Tomography problem/readout.}
The solver returns the trained weights only as amplitudes of a normalized
quantum state, so a dedicated readout stage is required, and its cost can
dominate the solver itself. Liu et al.\ analyze this step explicitly
via a two-stage sparse readout. First, the support of the
${r_{\rm out}}$ significant weights is identified by computational-basis sampling, a
weighted coupon-collector process requiring ${\mathcal{O}(r_{\rm out}\log r_{\rm out})}$
measurements, with a threshold caveat for vanishingly small weights. Second, the ${r_{\rm out}}$ amplitudes, including phases, are
recovered by a Clifford classical-shadow scheme at cost
${\mathcal{O}(m_{\rm out}^2r_{\rm out}^3/\epsilon_{\rm tom}^2)}$ in 1-norm error ${\epsilon_{\rm tom}}$,
plus polynomial classical post-processing \cite{Huang_2020, Liu_2024}. The construction is efficient only if the output sparsity ${r_{\rm out}}$ does not grow with the model size ${d}$. For dense outputs the same scheme recovers the ${r_{\rm out}}$ largest components, which the authors interpret as an intrinsic pruning step. Conversely, taken to the dense endpoint limit ${r_{\rm out}=d}$, the recovery cost becomes cubic in the output support, and even optimal full tomography requires ${\Omega(d/\epsilon_{\rm tom}^2)}$ copies \cite{Aaronson_2015}, so
full-vector readout would erase the solver's ${\mathrm{polylog}(d)}$
advantage. The output sparsity ${r_{\rm out}}$ and the readout tolerance therefore
replace the Hilbert-space dimension as the scaling variables.

\subsection{Dissipation and sparsity}
\label{sec:carleman_requirements}

\paragraph{Dissipation requirements}

Independently of the subsequent linear solver, the Carleman approximation is error-bound only under certain assumptions.
As higher order gradients beyond ${K}$ are truncated in Eq.~(\ref{eq:polynomial-update}), it needs to be ensured that the remaining gradient terms drive a parameter trajectory that does not diverge from the original one, i.e., the accumulated errors are well-controlled and bounded over the considered training interval.

The truncation error can be defined blockwise by $\varepsilon_j(t):=u^{\otimes j}(t)-\hat y_j(t),\ j=1,\ldots,{K}$, where $\hat y_j(t)$ is the $j$th block of the order-${K}$ truncated lifted system and $u^{\otimes j}(t)$ is the corresponding tensor power of the exact nonlinear trajectory, and $\varepsilon(t):=(\varepsilon_1(t),\ldots,\varepsilon_{{K}}(t))^T$.
Dissipativeness does not control the magnitude of the truncation terms themselves, but their propagation through the lifted dynamics. Following Liu et al., the linearization error satisfies $\varepsilon(t+1)=(I+A)\varepsilon(t)+\hat b(t)$, where $\hat b(t)$ contains the omitted tensor orders above the truncation boundary (\cite{Liu_2024}, II.19). The coefficients $F_\ell$ determine both the QCM $A$ and the boundary forcing $\hat b(t)$; however, these quantities play distinct roles. The bound on $\hat b(t)$ quantifies the error newly injected by truncation, whereas $I+A$ determines whether errors injected at previous steps decay or are amplified. If the QCM is fully dissipative, $\|I+A\|<1$, then the propagation is contractive and
\[
\|\varepsilon(T)\|
\leq
\frac{\max_t\|\hat b(t)\|}{1-\|I+A\|}.
\]
Thus, a truncation produces an error that remains bounded uniformly in the number of training steps.
Increasing the truncation order reduces the newly generated error while dissipativeness prevents its accumulation over time.

Full dissipation can be also phrased as the condition $|1+a_{\mu}|< 1$ where $a_{\mu}$ are the individual eigenvalues of the QCM, A. If all eigenvalues of $A$ are \textit{negative} and small ($-2<a_{\mu}<0$), the system is fully dissipative. We can further map this to the eigenvalues of the Hessian, assuming that this is the dominant order and $F_0=F_{\ell\geq 2}=0$ in the QCM:
The diagonal QCM block at tensor order $j$ is then a Kronecker sum of $j$ copies of $F_1=-\eta H$. It therefore has eigenvalues $a_{\alpha_1,\ldots,\alpha_j}=-\eta\sum_{i=1}^{j}h_{\alpha_i}$, where $h_\alpha$ are the Hessian eigenvalues. Under the given assumptions, dissipation requires $|1+a_{\alpha_1,\ldots,\alpha_j}|<1$, or equivalently $0<\eta\sum_{i=1}^{j}h_{\alpha_i}<2$. This is for example given, if all eigenvalues of $H$ are \textit{positive}, and the $\eta$ is sufficiently small.

In not fully dissipative systems, the truncation error can still be controlled over a finite time of steps $T$, e.g. by eigenvalue engineering, adjusting the truncation order or the learning rate. For simplicity, however, we will consider only fully dissipative systems in this article.

\paragraph{Sparsity requirements}

Sparsity plays a critical role at several distinct stages of the overall Carleman-linearized workflow to preserve the polylogarithmic quantum speedup:

\textbf{1) Gradient Calculation for the Carleman Matrix:}
% Explicitly constructing the Carleman system through degree $q$ from a dense model classically costs $O(qd^{q+1})$, which would erase any potential quantum advantage \cite{Liu_2024}. 
% Under certain requirements involving sparsity arguments, 
% one could argue that certain scenarios can relax this condition and enable a sublinear scaling. 
% We discuss the plausibility of such arguments in the results section.
For a dense model with $d$ parameters, explicitly constructing a Carleman system truncated at degree $q$ requires handling a lifted space of size $O(d^q)$. This scaling can only be avoided if the Carleman matrix is not constructed explicitly: instead, the original dynamics must be sparse, meaning that each update equation contains only a small number $S_{\max}$ of nonzero terms, and these terms must be generated on demand by an efficient sparse-access oracle. Under this assumption, the truncated QCM has row sparsity $O(qS_{\max})$ rather than $O(d^q)$ \cite{Liu_2024}(Theorem~V.1, Eq.~(V.98)). These assumptions are plausible for systems with local, bounded-degree interactions or compact rule-based descriptions, but they are strong for general NN: pruning weights alone does not guarantee efficient access to the remaining coefficients, and reading an arbitrary sparse model can still cost $\mathcal{O}(d)$ time. Hence, a potential quantum advantage additionally requires a compact oracle representation, sparse input and output weight vectors, sufficiently stable (dissipative) dynamics, and an efficient procedure for recovering the desired classical output.
%
%In the results section, we discuss the consequences of this step for the crossover estimation.

\textbf{2) State Preparation and Input Encoding:}
Uploading the initial sparse weight vector $\theta(0)$ to the quantum device would normally require Quantum Random Access Memory (QRAM), which is a challenging hardware assumption. Because $\theta(0)$ retains a bounded, sparse support $r$ independent of the model size $d$, it can instead be loaded efficiently using sparse-state-preparation circuits of depth $O(r \log d)$ without requiring QRAM \cite{Liu_2024,gleinig2022}.

\textbf{3) Matrix Oracle and HHL Simulation:}
The HHL algorithm's runtime scales polynomially with the row sparsity $s_M$ of the matrix oracle. By pruning the NN to reduce non-zero terms, the sparsity of the QCM $A$ is correspondingly bounded. This provides the critical sparse-access parameter $s_M$ required for the matrix oracle. If the QCM were dense, the query overhead would negate the exponential speedup in the matrix dimension.

\textbf{4) Tomography and Readout:}
The quantum solver outputs the trained weights as amplitudes of a quantum state. Extracting the full dense state would require $O(d/\epsilon_{\mathrm{tom}}^2)$ measurements, erasing the polylogarithmic advantage \cite{Liu_2024,aaronson2018shadow}. Instead, a two-stage sparse readout is used. First, the $r_{\rm out}$ most significant weights are identified via computational-basis sampling. Second, their amplitudes are recovered via a classical-shadow scheme with a cost scaling as $O(r_{\rm out}^3/\epsilon_{\mathrm{tom}}^2)$. This readout remains efficient only if the output sparsity $r_{\rm out}$ does not grow with the model size $d$, which intrinsically acts as a final pruning step.

\section{\uppercase{Method}}\label{sec:method}

\begin{figure}
    \centering
    \includegraphics[width=\linewidth]{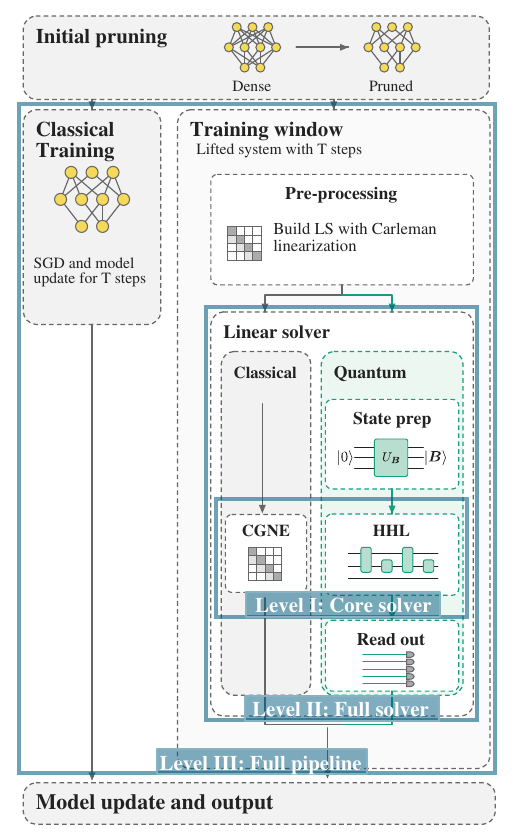}
    \caption{Overview of the different training approaches of a NN considered in this work, including training with SGD (left in gray), Carleman linearized approach from \cite{Liu_2024} with a classical CGNE solver (right in gray) and with a quantum HHL solver (right in green). Overview of the scope of different benchmarking levels (\texttt{Level I-III}) are marked in blue.}
    \label{fig:overview}
\end{figure}

%We estimate only a single training window of the pipeline of \cite{Liu_2024}, which covers $T$ training steps, and compare it to gradient decent (\cref{sec:classical_pipeline}), and on the quantum side, we consider the linearized pipeline of \cref{sec:quantum_pipeline}, which is shown in \cref{fig:pipeline_viz}. We show conventions, classical and quantum cost expressions, three comparisons, and the crossover dimension used in \cref{sec:results}.
We develop a three-level cost comparison between quantum and classical approaches as defined in \cref{fig:overview}, ordered by increasing scope and complexity. The cost computation for each component is described in \cref{sec:resource_estimation}. The crossover computation is given in \cref{sec:crossover}, and the setup of the entire pipeline is described in \cref{sec:experimental_setup}. The comparison is designed to make the bottlenecks of the pipeline more transparent as we can see how far a crossover point moves with increasing complexity. 

%\begin{figure}[h]
%    \input{figures/carleman_overview}
%    \caption{Visualization of the training pipelines proposed in~\cite{Liu_2024}. Standard SGD updates the weight vector $\theta$ sequentially at each time step $t$. The Carleman linearization approach instead enables multiple $t$s to be “overjumped” in a single operation, at the cost of solving an associated ODE, either classically or with a HHL-type solver. }
%    \label{fig:pipeline_viz}
%\end{figure}

\subsection{Resource estimates}\label{sec:resource_estimation}
To estimate an algorithm’s wall-clock execution time in seconds, we must determine the total number of relevant operations and either the time required to execute each operation or the corresponding execution rate. As summarized in \cref{tab:resources_estimate}, for classical algorithms, we consider Floating Point Operations (FLOPs) $N_F$ and the corresponding processing rate $r_{F}$ on the classical system, while for the quantum algorithm, we consider the number of $\mathcal{T}$-gates $N_{\mathcal{T}}$ within the quantum circuit and the gate times $\tau_{\mathcal{T}}$, since the $\mathcal{T}$-gate is the most expensive gate in universal fault-tolerant quantum computation.

\begin{table}[h]
    \centering
    \begin{tabular}{c|c c}
         &  Classical & Quantum \\ 
         \hline
         $\#$ of operations & $N_F$ (FLOPs) & $N_{\mathcal{T}}$ ($\mathcal{T}$-gate count)\\
         Time constants & $r_{F}$ (rate) & $\tau_{\mathcal{T}}$ (gate time) \\
         Time (seconds) & $N_{F}/r_{F}$ & $N_{\mathcal{T}}\tau_{\mathcal{T}}$
    \end{tabular}
    \caption{Estimation of the wall-clock execution time for classical and quantum algorithms. The classical time is obtained by dividing the total number of floating-point operations, $N_F$, by the classical processing rate, $r_F$, while the quantum time is estimated by multiplying the number of $\mathcal{T}$-gates, $N_{\mathcal{T}}$, by the time required to execute a single $\mathcal{T}$-gate, $\tau_{\mathcal{T}}$.}
    \label{tab:resources_estimate}
\end{table}
%\paragraph{Counts and times}
%The counts are independent of the used hardware and include the $\mathcal{T}$-gate count $N_{\mathcal{T}}$, the FLOP count $N_{\mathrm{F}}$, the number of one-sparse simulation calls $Q_{\mathrm{HS1}}$, and the number of state copies $n_{\mathrm{copies}}$. The (wall-clock) times, denoted by $t$, follow from these counts together with the hardware properties. On the quantum side one has $t=N_{\mathcal{T}}\tau_{\mathcal{T}}$, and on the classical side $t=N_{\mathrm{F}}/R_{\mathrm{cl}}$, as in \cite{tu2025identifyingpossiblefaulttolerantadvantage,beverland2022}.

\paragraph{Classical training (SGD).}
We estimate the cost of training a NN with $d$ parameters by stochastic gradient descent. 
As the operational execution costs in terms of FLOPs depend on the model architecture, we have to make further assumptions. Following \cite{Liu_2024}, we choose ResNet \cite{he2015deepresiduallearningimage} as a representative architecture, where the convolutional layers dominate the costs. Those include forward pass, backward pass, and the optimizer update. We denote by $\phi_{\mathrm{fwd}}$ the forward FLOPs per parameter and sample, and by $\gamma$ the ratio of the training FLOPs, forward and backward pass, to the forward FLOPs. The backward pass costs approximately twice as much as the forward pass, so we assume $\gamma=3$~\cite{epochai_backward_forward_2024}. The value of $\phi_{\mathrm{fwd}}$ for the reference network is determined empirically in Sec.~\ref{sec:experimental_setup}. Each parameter update adds around $2$ FLOPs. One window of $T$ steps with batch size $B$ thus requires 
\begin{align}
    N_{\mathrm{F}}^{\mathrm{SGD}}& = T\,d\,(\gamma\,\phi_{\mathrm{fwd}}\,B+2),
    %T_{Wall}^{SGD} &= T\,d\,(\gamma\,\phi_{\mathrm{fwd}}\,B+2)/r_F
    \label{eq:flops_sgd}
\end{align}
which is part of \texttt{Level III} benchmark (see \cref{fig:overview}).

\paragraph{Classical pre-processing (Carleman Linearization).}
Within one window, the solver propagates all $T$ steps from a single linearization point. The gradient that enters the expansion of the update is therefore evaluated only once per window, whereas SGD evaluates a gradient at each of the $T$ steps. One such gradient evaluation costs 
\begin{equation}
N_{\mathrm{F}}^{\mathrm{grad}}=\gamma\phi_{\mathrm{fwd}}d B
\label{eq:grad_flops}
\end{equation} FLOPs. Those costs have to be considered when solving the full linearized training pipeline.
Note that Eq.~\eqref{eq:flops_sgd} and Eq.~\eqref{eq:grad_flops} are the only points in our analysis, where explicit assumptions about the NN architecture enter. Our scaling study can be applied to any other model architecture, by adjusting those steps accordingly.

\paragraph{Classical LS (CGNE).}
Here, we estimate the cost of solving the linear system with classical Conjugate Gradient on the Normal Equations (CGNE), following the approach in \cite{tu2025identifyingpossiblefaulttolerantadvantage}. It solves the lifted system of dimension 
\begin{equation}\label{eq:history}
    N_h=(T+1)\sum_{j=1}^{K}d^{j}.
\end{equation}
One CGNE iteration performs two sparse matrix-vector products at $2N_hs_M$ FLOPs each, three vector summations at $2N_h$ FLOPs each, and $8N_h$ FLOPs for computing $\alpha$ and $\beta$, that is, $4N_hs_M+14N_h$ FLOPs in total \cite{tu2025identifyingpossiblefaulttolerantadvantage}. Following this, CGNE reaches the relative error $\epsilon_{\mathrm{CGNE}}$ after $(\bar\kappa_M/2)\ln(2/\epsilon_{\mathrm{CGNE}})$ iterations~\cite{tu2025identifyingpossiblefaulttolerantadvantage}. We set $\epsilon_{\mathrm{CGNE}}=\epsilon_{\mathrm{HHL}}$. Hence
\begin{align}
N_{\mathrm{F}}^{\mathrm{CGNE}} &=
\tfrac12\,\bar\kappa_M\ln(2/\epsilon_{\mathrm{CGNE}})\bigl(4s_M+14\bigr)N_h,\
%T_{Wall}^{CGNE} &= \tfrac12\,\bar\kappa_M\ln(2/\epsilon_{\mathrm{HHL}})\bigl (4s_M+14\bigr)N_h/r_F,
\label{eq:t_cgne}
\end{align}
where $N_h \propto d^K$ carries the dependency in $d$.
This cost is part of \texttt{Level I} and \texttt{Level II} benchmarks (see \cref{fig:overview}).

\paragraph{Quantum LS (HHL).} 
Here, we estimate the cost of solving the linear system with HHL following the approach in \cite{tu2025identifyingpossiblefaulttolerantadvantage}. It solves the lifted system of dimension \cref{eq:history} instead of performing classical training, similar to CGNE. To compute the number of required $\mathcal{T}$-gates, we follow the approach of \cite{tu2025identifyingpossiblefaulttolerantadvantage}. Firstly, we determine the number of 1-sparse Hamiltonian simulations (HS1) $Q_{\mathrm{HS1}}$ for HHL with sparsity $s_M$, condition number $\bar\kappa_M$ and required error rate $\epsilon_{\mathrm{HHL}}$ as

\begin{equation}
    Q_{\mathrm{HS1}}=\sqrt{\frac{320}{3}}\,
    \frac{\pi\bar\kappa_M^{2}s_M}{\epsilon_{\mathrm{HHL}}^{2}};
    \label{eq:q_hs1}
\end{equation}
We then multiply $Q_{\mathrm{HS1}}$ by the number of $\mathcal{T}$-gates per HS1 as \cite{tu2025identifyingpossiblefaulttolerantadvantage}
\begin{equation}\label{eq:N_QLS}
    N_{\mathcal{T}}^{\mathrm{QLS}} = Q_{\mathrm{HS1}} (18logN_{h}+90b_{\mathrm{prec}}+15), 
\end{equation}
which is part of \texttt{Level I-III}  benchmarks (see \cref{fig:overview}). Here, $b_{\mathrm{prec}}=7$ is the number of precision qubits. Since the history matrix is not Hermitian, the construction requires a Hermitian dilation.

\begin{equation}
\begin{aligned}
H(M_h)&=\begin{pmatrix}0&M_h\\M_h^\dagger&0\end{pmatrix},\\
N_H&=2N_h,\quad m_H=\lceil\log_2N_H\rceil,\quad N_{\mathrm{enc}}=2^{m_H}.
\end{aligned}
\label{eq:method_dimensions}
\end{equation}
The spectrum is $\pm\sigma_j(M_h)$, and the singular-value condition number is preserved. The compatible equation is $H(M_h)(0,x)^{\mathsf T}=(\vec b,0)^{\mathsf T}$ \cite{harrow2009}. We use later also the notation $m_h\approx K\log_2 d$.

%\MaxComment{The value of $Q_{\mathrm{HS1}}$ does not depend directly on $d$. The window length $T$ fixes $\bar\kappa_M$, while $r$ and $K$ fix $s_M$. The dimension $d$ enters through $m_h\approx K\log_2d$ below.}

\paragraph{Quantum input (state-preparation).}
Here, we estimate the cost of state preparation for HHL. Let $c_{\mathrm{qry}}$ denote the $\mathcal{T}$-gate cost per query. Then the oracle cost is
\begin{equation}
    N_{\mathcal{T}}^{\mathrm{orc}} = 2\,Q_{\mathrm{HS1}}\,c_{\mathrm{qry}},
    \label{eq:c_oracle}
\end{equation}
which is part of the \texttt{Level II-III} benchmarks together with the cost of the HHL solver from \cref{eq:N_QLS} (see \cref{fig:overview}) such as 
\begin{equation}
    N_{\mathcal{T}}^{\mathrm{run}}
    = N_{\mathcal{T}}^{\mathrm{QLS}} + N_{\mathcal{T}}^{\mathrm{orc}}.
    \label{eq:quantum_attempt_cost}
\end{equation}

\paragraph{Quantum readout (shadow tomography).}
Here, we estimate the cost of shadow tomography for HHL, which is part of the \texttt{Level II-III} benchmarks (see \cref{fig:overview}). Each measurement destroys the state, so every copy requires a new run. Only the first-order block at the last step of the solution vector holds the weight update, so $n_{\mathrm{copies}}$ counts accepted copies, that is, measurements that land in this terminal block, and the rejected ones are not priced. The readout follows the two steps of \cite{Liu_2024}. Support discovery measures in the computational basis until all $r$ nonzero weights have been observed, a coupon collector's problem with $\mathcal{O}(r\log r)$ measurements in expectation. The amplitudes are then recovered with the Clifford-shadow scheme of \cite{Liu_2024,Huang_2020}. Together, this requires
\begin{equation}
    n_{\mathrm{copies}}
    =\bigl\lceil r \ln r \bigr\rceil
    +\Bigl\lceil r^{3}/\epsilon_{\mathrm{tom}}^{2}\Bigr\rceil
    \label{eq:n_copies},
\end{equation}
which together with \cref{eq:quantum_attempt_cost} result in the cost of full quantum solver of \texttt{Level II-III} benchmarks as
\begin{equation}
   N_{\mathcal{T}}^{\mathrm{run}} \to n_{\mathrm{copies}} N_{\mathcal{T}}^{\mathrm{run}}.
    \label{eq:full_solver}
\end{equation}

%\MaxComment{For $T=100$ and $K=3$, the output state contains $101$ time blocks, each with three parts of dimensions $d$, $d^2$ and $d^3$. To update the model parameter, only the states representing $d=1$ and $T=100$ are relevant. So rejected measurements outside this part require a new preparation of the state and we assume accessing this state can be done efficiently and there are routines to improve the probability \cite{Liu_2021}.For the ideal normalized state, the acceptance probability is
%\begin{equation}
%p_{\mathrm{target}}=
%\frac{\|\hat y_1(T)\|^2}{\sum_{t=0}^{T}\|\hat y(t)\|^2}.
%\label{eq:target_selection}
%\end{equation}}

% A run succeeds only when the HHL postselection succeeds. Its probability satisfies $p_{\mathrm{HHL}}\ge1/\bar\kappa_M^{2}$~\cite{harrow2009}, this is not considered \cite{tu2025identifyingpossiblefaulttolerantadvantage}. Amplitude amplification requires $a_{\mathrm{HHL}}=\bar\kappa_M$ coherent rounds. We set $a_{\mathrm{HHL}}=1$ here, and we discuss the effect of amplification in \cref{sec:results}.

\subsection{Solution pipelines scalings}
\label{sec:crossover}

In \texttt{Level I}, we only compare the core linear solvers in both the classical and the quantum pipelines, see Fig.~\ref{fig:overview}.
On the quantum side, this means solving an already prepared linear equation in a single run.
By Eq.~\eqref{eq:q_hs1}, $Q_{\mathrm{HS1}}$ is proportional to $s_M$, since the condition number $\bar\kappa_M=2(T+1)$ and the precision $\epsilon_{\mathrm{HHL}}$ do not depend on $r$ or $d$ and hence enter only as a constant \cite{tu2025identifyingpossiblefaulttolerantadvantage} where $Q_{\mathrm{HS1}}\propto\kappa^{2}s/\epsilon^{2}$. For large $r$, $s_M\approx Kr^{2}$, so that the sparsity grows quadratically with $r$. The history register has $m_h\approx K\log_2 d$ qubits and thus contributes only a logarithmic factor.
Therefore, the core solver requires a wall-clock time $t_{\mathrm{Q,I}}(d,r)$ scaling as (with Eq.~(\ref{eq:N_QLS})),
\begin{align}
t_{\mathrm{Q,I}}(d,r)
= N_T^{\mathrm{QLS}}(d,r)\tau_T 
= \mathcal{O}\!\left(r^2\log d\right).
\end{align}
The classical reference on this level is the CGNE linear solver from Eq.~(\ref{eq:t_cgne}),
\begin{align}
\label{eq:lvlIandII_cls}
t_{\mathrm{CGNE}}(d,r)
= \frac{N_F^{\mathrm{CGNE}}(d,r)}{r_F}
= \mathcal{O}\!\left(r^2d^K\right).
\end{align}
On this level, as expected \cite{tu2025identifyingpossiblefaulttolerantadvantage}, the HHL solver scales favorably compared to the classical baseline.\\

For \texttt{Level II}, we include additional steps the quantum solver requires in a more realistic picture. In particular, this concerns the transformation interfaces between the quantum state and the classical world, represented by the state preparation of the HHL input and the quantum tomography readout which requires running various copies of the core solver. While the state preparation can be done efficiently in $d,r$ (see Eq.~\eqref{eq:quantum_attempt_cost}), the readout changes the scaling in $r$. For large $r$, the tomography term in Eq.~\eqref{eq:n_copies} dominates and one has $n_{\mathrm{copies}} \propto r^{3}$ copies.
The full quantum solver then achieves a wall-clock time $t_{\mathrm{Q,II}}(d,r)$ of (see Eq.~(\ref{eq:full_solver})),
\begin{align}\label{eq:level_II}
t_{\mathrm{Q,II}}(d,r)
&= n_{\mathrm{copies}}(r)N_T^{\mathrm{run}}(d,r)\tau_T
= \mathcal{O}\!\left(r^5\log d\right).
\end{align}
The classical comparison is the same as in Eq.~\eqref{eq:lvlIandII_cls}.
The quantum window cost therefore still has a favorable scaling in $d$, compared to the classical baseline, but the $r^{5}$ creates harsher constraints on the sparsity.
If one assumes a sparsity that is a power fraction of the model parameters itself -- meaning $r=d^{\beta}$ (see Sec.~\ref{sec:results}) -- a upper bound of $\beta < \frac{1}{5}$ is identified to protect the scaling in $d$. \\

At \texttt{Level III} we compare the entire training pipeline.
Most importantly, the gradients that form the linearized Carleman matrix have to be derived first, with costs proportional to $d$, see Eq.~\eqref{eq:grad_flops}. 
The quantum-supported linearized training pipeline then scales as
\begin{align}
t_{Q,III}(d,r)
= \frac{N_F^{\mathrm{grad}}(d)}{r_F}
  + t_{\mathrm{Q,II}}(d,r)
= \mathcal{O}\!\left(d+r^5\log d\right).
\end{align}
In comparison, the classical SGD baseline from Eq.~\eqref{eq:flops_sgd} reads for \texttt{Level III}
\begin{equation}
t_{\mathrm{SGD}}(d)
= \frac{N_F^{\mathrm{SGD}}(d)}{r_F}
= \mathcal{O}(d).
\end{equation}
The gradient preparation step ultimately leads to the same linear scaling in $d$ for both the quantum and the classical pipeline, thus erasing any large-scale quantum advantage. We discuss in Sec.~\ref{sec:discussion}, in which scenarios the cost of gradient preparation might be alleviated and how intermediate crossover regimes might still be relevant for possible quantum speedups. The quantitative crossovers for the described concepts are derived in Sec.~\ref{sec:results}.

\subsection{Sparsity scaling}
\label{sec:sparsity_scaling}
Since sparsity is one of the key assumptions of this approach (see Sec.~\ref{sec:carleman_requirements}), each benchmarking scenario (\texttt{Level I--III}) is evaluated under different assumptions about the system's sparsity $r$. These range from a constant sparsity that is independent of the model size -- such as $r=100$, which constitutes a highly optimistic assumption -- to more realistic scaling models in which the sparsity depends on the number of NN parameters $d$. Examples include pruning the NN to $10\%$ of its size, corresponding to $r=0.1d$, as well as alternative scaling laws such as $r=\log_2(d)$ and $r=d^{0.1}$.

\subsection{Experimental setup}\label{sec:experimental_setup}
\paragraph{Neural Network setup.}
Our representative architecture is a ResNet model \cite{he2015deepresiduallearningimage}, where most of the compute is attributed to convolutional layers. 
%with $d=7{,}451{,}044$ parameters
For the cost analysis of Eqs.~\eqref{eq:flops_sgd} and \eqref{eq:grad_flops} we use
batch size $B=2048$ and a time horizon of $T=100$ for this benchmark. A dense layer yields $\phi_{\mathrm{fwd}}=2$, whereas a convolutional layer has a larger value, because each kernel weight is reused at every output position. 
The reference ResNet has depth 32 and widen factor 4~\cite{Liu_2024}. Counting the FLOPs of its matrix products and convolutions for $32\times32$ inputs gives $\phi_{\mathrm{fwd}}\approx296$. We find this value to be empirically stable also under architecture variations with different amounts of layers.

\paragraph{Classical computational rate.}
We set the classical computational rate to $r_{F}=10^{9}$ operations per second \cite{tu2025identifyingpossiblefaulttolerantadvantage}. Counting FLOPs makes the classical cost independent of the hardware, but it does not capture memory access or hardware utilization~\cite{bartoldson2023compute}. The classical times are therefore estimates of the arithmetic work and not measured runtimes. %\AlonaComment{the memory wall problem is the growing performance gap between rapidly improving CPU/GPU compute speeds and much slower improvements in memory bandwidth, causing processors to stall while waiting for data}

\paragraph{Quantum gate time.}
%We set the logical $\mathcal{T}$-gate time to $\tau_{\mathcal{T}}=c_{\mathrm{eff}}\tau_{\mathrm{cyc}}=110\,\mathrm{ns}$. 
We assume that distilled magic states are available on demand and neglect their preparation time. The logical $\mathcal{T}$-gate time is set to $\tau_{\mathcal{T}}=10\,\mathrm{ns}$, corresponding to a single cycle \cite{tu2025identifyingpossiblefaulttolerantadvantage} which is the time required to implement the gate using a distilled magic state. 
%\AlonaComment{double check}

\paragraph{HHL setup.}
% For the precision $\epsilon_{\mathrm{HHL}}=0.01$, we set $b_{\mathrm{prec}}=\lceil\log_2(1/\epsilon_{\mathrm{HHL}})\rceil=7$, and the query cost is $c_{\mathrm{qry}}=1$.

For the precisions $\epsilon_{\mathrm{HHL}}=\epsilon_{\mathrm{tom}}=0.01$. Further, we set $b_{\mathrm{prec}}=7$, and the query cost is $c_{\mathrm{qry}}=1$.

\paragraph{The cost model.} We assume that the coefficient tensors $F_\ell$ can be computed analytically up to truncation order $K=3$ and that the weight vector remains sparse \cite{Liu_2024}. 
%The term $F_0$ still has to be computed once per window. 
We grant dissipativity instead of verifying it, so that the pipeline is priced conditional on $\lVert I+A\rVert<1$, which can be observed empirically in early training but is not guaranteed at every stage of training or for an arbitrary model \cite{Liu_2024}. From this assumption follows the condition-number bound $\kappa_M\le\bar\kappa_M=2(T+1)$ \cite{Liu_2024}. As in \cite{tu2025identifyingpossiblefaulttolerantadvantage}, the oracle supplies the history matrix as a sum of $s_M$ one-sparse terms.

The quantum window returns $r$ normalized amplitudes of the weight update, while SGD returns all $d$ weights, so \texttt{Level III} compares neither equal outputs nor equal model quality. Both pipelines use the same dense gradient.

\section{\uppercase{Results}}\label{sec:results}
%For $r=4$, one run needs $1.7\times10^{15}$ $\mathcal{T}$ gates, at $\tau_{\mathcal{T}}=110$\,ns that is $1.9\times10^{8}$ seconds. The readout needs $6{,}406$ fresh runs, so the quantum window costs $1.2\times10^{12}$\,s, while $T$ steps of dense SGD at $10^{9}$\,FLOP/s  cost $1.4\times10^{6}$\,s. At the ResNet's models's size the modelled quantum cost is about $9\times10^{5}$ times higher.

% Exact wall-clock model used in the three panels; all times in seconds.

% All parameters except d and r are held fixed; d >= 2 and r >= 1.

\begin{figure*}[h]
    \centering
    \includegraphics[width=\textwidth]{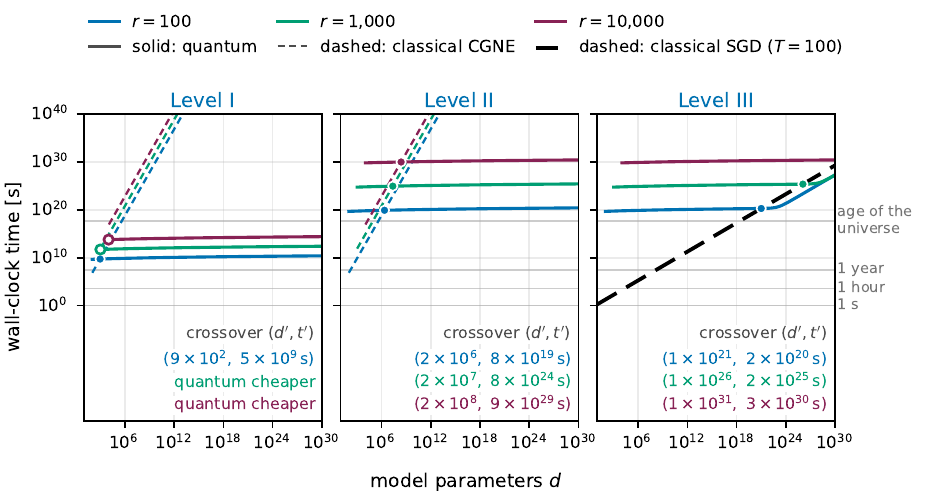}
    \caption{\texttt{Level I - III} for the constant sparsities $r=100$ (blue), $r=1{,}000$ (green) and $r=10{,}000$ (purple) at $K=3$, $T=100$, $r_F=10^{9}\,\mathrm{FLOP/s}$, $\tau_{\mathcal{T}}=10\,\mathrm{ns}$. Solid curves are the quantum estimation, dashed curves the classical one, the colour gives $r$, and a lower curve means a faster runtime. \texttt{Level I} (core solver): HHL vs CGNE solve of the lifted history system $N_h$. The core grows only as $\log d$, CGNE as $d^{3}$, and both scale with $s_M$, so the crossover point $d'\approx10^{3}$ for $r=100$, while for $r=1{,}000$ and $r=10{,}000$ the quantum core is already cheaper at the smallest admissible size $d=r$. \texttt{Level II} (full solver): HHL with state preparation and tomography vs CGNE solve. Every copy repeats the run and $n_{\mathrm{copies}}$ grows as $r^{3}$, so $d'$ moves with $r$. \texttt{Level III} (full pipeline): HHL with state preparation and tomography plus classical preprocessing vs $T=100$ steps of SGD (black). The quantum curves stay nearly flat until the one classical gradient calculation per window dominates.}
    \label{fig:comparisons-constant}
\end{figure*}
\begin{figure*}[t]
    \centering
    \includegraphics[width=\textwidth]{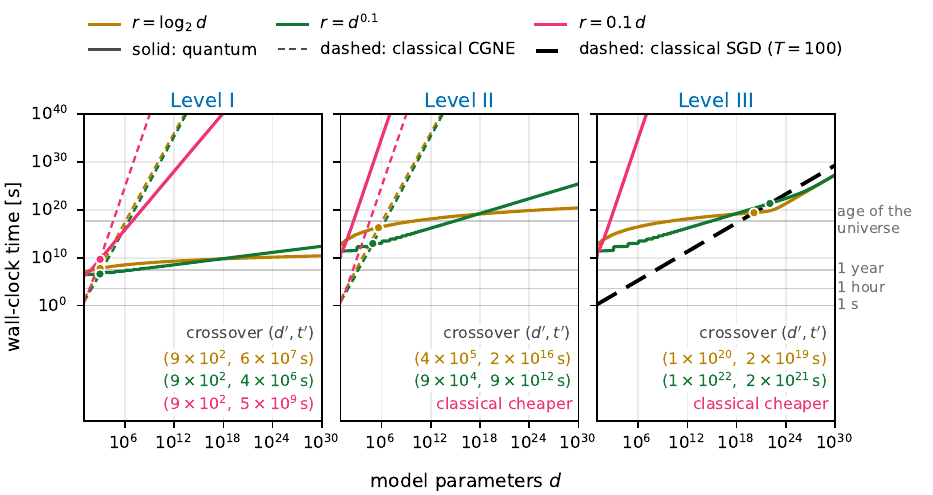}
    \caption{\texttt{Level I - III} for sparsity laws that grow with the model, $r=\log d$ and $r=d^{0.1}$, and the fixed fraction $r=0.1\,d$ as reference, at $K=3$, $T=100$, $r_F=10^{9}\,\mathrm{FLOP/s}$, $\tau_{\mathcal{T}}=10\,\mathrm{ns}$. Line styles, markers and exclusions as in Fig.~\ref{fig:comparisons-constant}. In the right panel, a crossover requires $r$ to grow slower than $d^{1/5}$. The steps of the $d^{0.1}$ curve come from the integer support $\lceil r\rceil$ and the integer copy counts.}
    \label{fig:comparisons-sublinear}
\end{figure*}

\subsection{Realistic resource estimation}

\paragraph{Level I: Solver Core}
Figure~\ref{fig:comparisons-constant}(left) isolates the quantum linear-system solver core against the classical CGNE algorithm applied to the lifted system of dimension $N_h$. For $r=100$, the HS1 solver achieves a crossover with CGNE at approximately $d' \approx 10^3$ parameters and $t
' \approx 10^9$ seconds. This crossover reflects the fundamental algorithmic advantage: the quantum solver scales as $\log d$ compared to CGNE's $d^3$ scaling. Both exhibit proportional dependence on the matrix sparsity $s_M$, but the exponential gap in dimension dominance emerges for sufficiently large systems. At $r = 10^3$ and $r = 10^4$, the crossover occurs at the minimal admissible dimension $d = r$, indicating that sparse-support constraints become the limiting factor rather than system dimension.

\paragraph{Level II: Full Solver}
When accounting for repeated state preparation and destructive measurement (Figure~\ref{fig:comparisons-constant}(middle)), the quantum advantage erodes significantly. Here, $n_{\mathrm{copies}}$ grows as $r^3$ due to tomography requirements (Eq.~\eqref{eq:n_copies}), while the per-run cost remains proportional to $r^2 \log d$. The combined cost thus scales as $r^5 \log d$. The crossover point $d'$ now depends strongly on $r$: for $r=100$ it occurs near $d' \approx 10^6$ and $t' \approx 10^{19}$ seconds, while for $r = 10^4$ it shifts to $d' \approx 10^8$ and $t' \approx 10^{29}$ seconds. This $r$-dependence reveals that readout becomes a severe bottleneck, amplifying the challenge of extracting quantum advantage. The readout overhead dominates the solver cost by multiple orders of magnitude.

\paragraph{Level III: Full Pipeline}
The most realistic comparison (Figure~\ref{fig:comparisons-constant}(right)) adds one classical gradient evaluation per window to the quantum costs and compares the result against $T = 100$ steps of standard SGD on a CPU. Under the assumption of constant sparsity ($r = 100$), the crossover moves to $d' \approx 10^{21}$.
While the sparsity constraints in \texttt{Level I/II} shift the crossover points towards larger model and time horizons, they do not change the asymptotic scaling in $d$. 
Classical pre-processing in the full \texttt{Level III} pipeline, however, adds a linear dependency in $d$ to the workflow (same as the classical SGD), which originates from the need to prepare the Carleman matrix with gradients calculated from the loss function for each parameter. 
This reduces the possible quantum advantage in this asymptotic regime to a mere offset. In this scenario, only an intermediate crossover regime is left as an interesting window of practical speedups. 
% quantum window cost plateaus at approximately $1.2 \times 10^{12}$ seconds across all model sizes shown, because the readout dominates and does not depend significantly on $d$. 
However, this crossover occurs at compute times exceeding $10^{20}$ seconds ($\mathcal{O}(10^{13})$ years), well beyond any practical training regime. 

\paragraph{Summary}
The central observation is that practical quantum advantage exists only under highly restrictive conditions. \texttt{Level I} shows that the solver core alone can outperform classical linear solvers at modest system sizes. However, once realistic readout costs are included (\texttt{Level II}), the required model size increases by more than three orders of magnitude. Finally, when the full training pipeline is considered (\texttt{Level III}), achieving advantage requires either extremely sparse outputs, extremely large models ($d \gtrsim 10^{20}$), or extremely long training horizons ($> 10^{19}\,\mathrm{s}$). These conditions are mutually impractical for NN training. However, note that this analysis assumes sequential processing; parallelization would significantly affect the estimated crossover, as discussed in \cref{sec:discussion}.

\subsection{Effect of sparsity scaling}

\cref{fig:comparisons-constant} assumes that sparsity remains constant as the system dimension $d$ increases. This is a strong assumption, so in \cref{fig:comparisons-sublinear} we investigate sparsity functions that depend on $d$. The results show that a sparsity of $r = 0.1d$ is too demanding for HHL as under this scaling, neither the \texttt{Level II} nor the \texttt{Level III} comparison exhibits a crossover. Unlike the 10\% scaling law, the other scaling laws exhibit similar behavior to one another, with crossover points at comparable positions. 
Their crossover points are more favorable than those in \cref{fig:comparisons-constant}, in particular for the \texttt{Level II} benchmark. 

% \newpage
% \begin{figure*}[!htb]
%     \centering
%     \includegraphics[width=0.8\textwidth]{figures/absolute_classical_time.png}
%     \caption{Absolute classical arithmetic-time proxies for $K=3$ at $1\,\mathrm{GFLOP/s}$. The left panel shows the conjugate-gradient estimate, while the right panel shows dense stochastic gradient descent. The color scale gives the base-10 logarithm of the estimated runtime in seconds.}
%     \label{fig:absolute-classical-time}
% \end{figure*}

% \subsection{Crossover point}

% HS1 T gates and oracle queries
% state-preparation depth and classical construction work
% fresh solution-state copies for support discovery and tomography
% shadow-Clifford work
% expected solve attempts under selectable postselection/amplification

\newcommand{\SubItem}[1]{
    {\setlength\itemindent{15pt} \item[-] #1}
}

\section{\uppercase{Discussion}}\label{sec:discussion}
%The results of this study are sobering. 
While many previous works make strong assumptions and predict substantial asymptotic improvements over classical routines, we constructed a complete end-to-end pipeline and evaluated its concrete performance. For the pipeline considered here, the current estimates indicate that the crossover point lies beyond practical limits. Nevertheless, these results provide a needed benchmark for future quantum software and hardware development in order to bring the crossover point within reach.

\paragraph{Bottlenecks.}
Within the described approach, we identify the two main obstacles for a practical quantum advantage.

First, the full quantum-supported training pipeline at \texttt{Level III} requires access to the polynomial coefficient tensors $F_\ell$, which means calculating the loss function gradients for each parameter. This incurs a cost of order $\mathcal{O}(d)$ that erases the favorable scaling of the HHL core solver.
This overhead could only be avoided if the update rule admits a compact, structured representation, for example, through locality, sparsity, parameter sharing, symmetries, or an analytically specified polynomial dynamics, that permits the required nonzero coefficients to be generated on demand by the matrix oracle. In such settings, gradients and higher-order coefficients need not be explicitly derived or stored as dense tensors. For generic dense NN losses, however, no such construction is available. A possible silver lining might be in intermediate regimes. Our findings in Sec.~\ref{sec:results} suggest that -- accepting the long-tail linear-in-$d$ scaling at \texttt{Level III} -- quantum support can still provide relevant quantitative speed-ups. For example, we see in Fig.~\ref{fig:comparisons-constant} that just beyond the crossover point $d'$ follows a model size window with the maximum positive gap in wall-clock time. Depending on the configuration setting and sparsity, this regime may well save a few orders of magnitude in compute time, which is an opportunity in itself.

Second, a growing number of non-zero entries $r$ shifts the cross-over points to practically infeasible regimes, see Sec.~\ref{sec:results}. Choosing small $r$, however, defies the purpose of training large-scale models, as it means that a vast majority of trained network parameters are zero. The cost in $r$ is largely driven by the tomography step ($\sim r^3$). The approach therefore demands more efficient tomography methods, or alternatively, the classical pipeline around the quantum LS solver could be redesigned to process partial solutions, potentially eliminating the need for tomography.

%---------------------------------------------
%\paragraph{Required logical gate time improvements}
%The execution time of quantum logical operations must be reduced through improvements in the underlying hardware, which will require advances throughout the software stack, including controllers, transpilers, compilers, and, in particular, quantum error correction. 
%\AlonaComment{Magic state distillation time requirements}
%\AlonaComment{Add overview of MQSS developments, discuss the effect of QEC on logical gate time}
%\AlonaComment{Impact of decoder latency \cite{khalid2026impactsdecoderlatencyutilityscale}}
%\AlonaComment{NVIDIA approach to QPU control speed up \cite{caldwell2025platformarchitecturetightcoupling}}

%---------------------------------------------
\paragraph{Limitations and future work.} 
Our analysis comes with a number of conceptual limitations, which should be stated explicitly.
\begin{itemize}
    \item \textbf{ML training pipeline:} We focus on the proposed ML training pipeline from \cite{Liu_2024} and therefore inherit its limitations. For example, the assumption of a fully dissipative model, required for the error and condition-number bounds used here, is highly restrictive. In this setting, the loss Hessian cannot have any negative or zero eigenvalues, an unnatural requirement for typical NNs that necessitates special preparation. 
    Additionally, the linearized pipeline relies on theoretically derived error bounds for parameter trajectories, but even small deviations in these trajectories could possibly imply significant differences in the trained network's performance. It is out of scope for this work to evaluate the accuracy of models whose parameters are trained using the different pipelines. To add to that, we generally assume a training window of $T$ time steps. Generally, this will not be sufficient for a realistic training pipeline, which is why \cite{Liu_2024} describes a step-wise approach, where multiple Carleman bursts are performed in sequence, each with additional pre- and post-processing. A possible speedup is expected only within a Carleman window.  
    \item \textbf{Classical and quantum baselines:} In this work, we focused on specific algorithms for cost estimation. However, CGNE may not be the most efficient classical solver for general linear systems \cite{rude2026cg}. Similarly, other QLS may offer more favorable scaling or relax demanding assumptions about sparsity or error bounds \cite{dervovic2018quantumlinearsystemsalgorithms, Gily_n_2019}, as well as other tomography methods 
    \cite{Huang_2020, Schmale_2022} may be more resource-efficient. In future work, we plan to investigate alternative baselines and assess how they affect the crossover point.
    \item \textbf{Cost model abstractions:} To avoid excessive complexity, we make several abstractions. On the classical side, our cost model omits memory wall problem and hence intermediate operations such as memory allocation, data transfer, and data writing, as well as bandwidth bottlenecks. Additionally, we considered hypothetical NN sparsity functions. Investigating how sparsity scales in practice would help contextualize our findings. On the quantum side, it abstracts away magic state distillation, the effects of the specific QEC method, and differences between quantum hardware (e.g. between superconducting and ion-trap qubits). Additionally, the analysis counts only accepted copies in the terminal block. The fraction of accepted shots can be increased by appending idle steps to the linear system, which also increases the condition number, or by amplitude amplification~\cite{berry2014highorder,Liu_2021}. These steps, the rejected shots and the repetitions needed for the success probability of the HHL solver itself are not priced. Future work could incorporate these omitted steps to provide even more realistic crossover estimates.
    \item \textbf{Cost model parameters:} We use estimates from \cite{tu2025identifyingpossiblefaulttolerantadvantage}, which assume a relatively pessimistic CPU speed ($10^9$ FLOP/s) and an optimistic QPU speed ($10$ ns per cycle), even for superconducting hardware. Moreover, the CPU benchmark is not representative of typical real-world applications. We discuss parallelization below and leave the development of more realistic cost estimates to future work.
%    \item \textbf{Cost model not fully end-to-end:} Intermediate steps such as memory allocation, data transfer and writing, or bandwidth bottlenecks are not priced in our simple cost model.
\end{itemize}

%For the classical computation, we deliberately assume a conservative performance of $10^9\,\mathrm{FLOP/s}$ following \cite{tu2025identifyingpossiblefaulttolerantadvantage}. Modern GPUs can achieve approximately $10^{12}\text{--}10^{14}\,\mathrm{FLOP/s}$, which would shift the crossover point even further beyond the practically feasible regime.

%---------------------------------------------
\paragraph{Parallelization.}
The reported crossover times seem prohibitively large, however, they must be interpreted as CPU estimates rather than as GPU or HPC\footnote{High Performance Computing (HPC)} wall-clock predictions. Efficient parallelization could drastically reduce the wall-clock time.

Such techniques are well established on the classical side.
Classical SGD can exploit data, tensor/model, and pipeline parallelism, allowing GPU clusters to convert large aggregate computational costs into substantially shorter wall-clock times, subject to communication, synchronization, and load-balancing overheads. 
The large-scale LLM training runs discussed in the introduction illustrates it: the approximately $10^6$ GPU-hours used for LLaMA-65B correspond to about 21 days of elapsed time when distributed over hundreds of GPUs, a reduction in wall clock time of more than three orders of magnitude. With similar methods, the crossover times reported in this study could be reduced drastically.

On the quantum side, such a parallelization of quantum algorithms and distributed QC is an active research field \cite{Zhang2024parallelquantum, Shayegan2025-rw, Main2025-pn}. However, its practical application particularly for QLS remains to be demonstrated. These challenges span several layers. Architecturally, modular fault-tolerant designs must support cross-module error correction and synchronization. At the hardware level, low-latency quantum interconnects are required. At the logical level, distributed logical operations and sufficiently high magic-state throughput are needed. At the software level, compilers and runtimes must be able to partition and schedule work across modules. Finally, the resulting resource overheads must not offset the gains from parallelization. Future research should focus on strongly parallelized implementations of QLSs, which could realize scaling improvements analogous to those that have enabled the sustained progress of classical computing and shift the crossover point into a more practically relevant regime.

%\paragraph{Future extensions.}
%\AlonaComment{Add quantum speed limits as the idealistic lower benchmark}
%\AlonaComment{Add discussion on extending to other NN architectures, e.g. LLMs, and considering more realistic sparsity scaling}

% depdency on qubits

% error mitigation assumptions

% FTQC assumptions
%\subfile{sections/limitations.tex}

\section{\uppercase{Conclusion}}
\label{sec:conclusion}
In this work, we present an end-to-end resource estimate for a quantum-enhanced ML training pipeline \cite{Liu_2024} with the resource model of \cite{tu2025identifyingpossiblefaulttolerantadvantage} at three comparison levels. Our results indicate that at the solver core level (QLS vs classical ODE comparison) quantum-enhanced pipeline can outperform classical counterpart pipeline at approximately $\sim 10^3$ model size. That result does not, by itself, imply a training advantage, as the first comparison level neglects the cost of the rest of the pipeline. At the full pipeline level (entire preprocessing and QLS vs SGD) for a reference ResNet model, the crossover is estimated at $> 10^{20}$ model size and $> 10^{19}$ seconds of computation time and is highly dependent on the sparsity of the model.

More broadly, this work provides a blueprint for evaluating quantum-enhanced ML algorithms under realistic hardware constraints. By making the full cost structure of the pipeline explicit, it identifies the specific engineering and algorithmic advances, such as lightweight input and output subroutines, more efficient tomography, structured gradient representations, and parallelization in quantum algorithms, that must be realized to move the crossover point into a practically relevant regime. We hope these results serve as a quantitative benchmark against which future quantum software and hardware development can be measured.

%The solver core and repeated state production for destructive readout are compared with CGNE on the lifted system. The window, which additionally includes one classical gradient, is compared with dense SGD. For $T=100$, $K=3$ and $r=100$, the solver core intersects CGNE near $d=9.3\times10^2$ and does not indicate a training advantage.
%Once repeated state production and one classical gradient per window are included, the selection-free quantum run costs about $6.0\times10^{13}$ times the dense-SGD estimate at the reference size of a default ResNet $(d\approx7\times 10^6$ parameters). The intersection with SGD occurs only near $d=1.0\times10^{21}$, at the deliberately slow classical rate $r_F=10^9$\,FLOP/s.
%Furthermore, we have shown how important constant or slowly growing sparsity is, since with a $10\%$ retained support, no intersection with SGD is found.
%Constructing higher order $F_{\ell}$ and terminal selection remain unpriced in the reported comparisons. Further circuit and interface costs are also omitted, while the outputs and training quality are not matched. Such training conditions are not guaranteed for an arbitrary model architecture and remain unverified for the reference network. Establishing an operational crossover therefore still requires complete cost accounting, justified dynamical and readout assumptions, and a common performance target. The present results identify conditional cost intersections for an idealised family of training windows. They do not demonstrate lower cost for neural-network training as a whole.

\section*{\uppercase{Acknowledgements}}
The research is part of the Munich Quantum Valley, which is supported by the Bavarian state government with funds from the Hightech Agenda Bayern Plus.

% \todo[inline]{
% \begin{enumerate}
% \item \sout{GD Flop estimation per parameter per training sample for ResNet (Florian)}
% \item Add results about scaling of quantum classical: 1) Tu paper (HHL for perfect in/out-ode vs CG), adjustments with Eisert terminology
% 2) full solve ODE (HHL vs CG)
% 3) compare pipline gradient descent vs linearized ode with HHL
% (Max)
% \item Homogenize notation in 4.1
% \item \sout{Parameter overview and setup (Max)}
% \item Background (Sec. 3): oracle problem (Andrea) 
% \item \sout{Related work (Sec. 2), e.g. Tu Paper (Max)}
% \item Discuss assumptions of sparsity (Sec. 3 or 4)
% \item Discuss assumption of classical scaling parts (e.g. gradient calculation for carleman matrix)
% \item \sout{Sec 4: Assumption, Scaling, Crossover point, Evaluation, structure by 1), 2), 3) - create structure of Sec. 4 (Florian)}
% \item Further extension (non-dissipative model, realistic initialization, encoding, read-out, etc.)
% \end{enumerate}
% }

\printbibliography

\begin{appendix}

\end{appendix}

\end{document}